\documentclass[prb, reprint, amsmath,amssymb,aps]{revtex4-1}
\usepackage{graphicx}
\usepackage{dcolumn}
\usepackage{bm}
\usepackage{amsmath}
\newcommand{\ks} {{\bf k}}
\newcommand{\es} {{\bf e}}
\newcommand{\ac} {{\bf A}}
\newcommand{\kap} {{\boldsymbol \kappa}}

\begin{document}

\title{Ultrafast photoionization and excitation of surface-plasmon-polaritons on diamond surfaces }

\author{Tzveta Apostolova$^{1,2}$}

\author{B. D.  Obreshkov$^1$}

\affiliation{$^1$ Institute for Nuclear Research and Nuclear Energy, 1784 Sofia, Bulgaria}
\affiliation{$^2$ Institute for Advanced Physical Studies, New Bulgarian University, 1618 Sofia, Bulgaria}

\author{
A.A. Ionin$^3$, S.I. Kudryashov$^{3,4}$
S.V. Makarov$^{3,4}$ N.N. Mel'nik$^3$ A.A. Rudenko$^3$}

\affiliation{ $^3$Lebedev Physical Institute, 119991 Moscow,
Russia}

\affiliation{ $^4$ ITMO University, 197101 St. Petersburg,
Russia}

\begin{abstract}

Ultrafast plasmonics of novel materials has emerged as a promising
field of nanophotonics bringing new concepts for advanced optical
applications. Ultrafast electronic photoexcitation of a diamond
surface and subsequent surface plasmon-polaritons (SPPs)
excitation are studied both theoretically and experimentally - for
the first time. After photoexcitation on the rising edge of the
pulse, transient surface metallization was found to occur for
laser intensity near 18 TW/cm$^2$ due to enhancement of the impact
ionization rate; in this regime, the dielectric constant of the
photoexcited diamond becomes negative in the trailing edge of the
pulse thereby increasing the efficacy with which surface roughness
leads to inhomogeneous energy absorption at the SPP wave-vector.
These transient SPP waves imprint permanent fine and coarse
surface ripples oriented perpendicularly to the laser
polarization. The theoretical modeling is supported by the
experiments on the generation of laser-induced periodic surface
structure on diamond surface with normally incident 515-nm, 200-fs
laser pulses. Sub-wavelength ($\Lambda \approx 100$ nm) and near
wavelength ($\Lambda \approx 450$ nm) surface ripples oriented
perpendicularly to the laser polarization emerged within the
ablative craters  with the increased number of laser shots;
the spatial periods of the surface ripples decreased moderately with the
increasing exposure. The
comparison between experimental data and theoretical predictions
demonstrates the role of transient changes of the dielectric
permittivity of diamond during the initial stage of periodic
surface ripple formation upon irradiation with ultrashort laser
pulses.

\end{abstract}

\maketitle

\section{Introduction}

Diamond is a material, exhibiting unique mechanical, thermal and
electrical properties, as well as high electron and hole mobility
\cite{Isberg2002}, promoting its high performance in
microelectronic devices. At the same time, diamond is a basic
ingredient in modern nanophotonics
\cite{Aharonovich2014,Hausmann2012}. Due to its high refractive
index in UV-VIS range, it is prospective material for
all-dielectric \cite{Kuznetsov2016}  and even hybrid
metal-dielectric nano-photonic devices and circuits
\cite{Ozkan1999,Pham2016,Ivanova2016}. Moreover, despite its
dielectric character, similarly to silicon it can be promptly
turned by intense ultrashort laser pulses into short-lived
plasmonic state, becoming so-called "virtual plasmonic material",
supporting photoexcitation and propagation of surface
plasmon-polaritons (SPPs)
\cite{Boltasseva2011,Kumar2013,Danilov2015}, for potential
applications in ultrafast optical switching, spatial phase
modulation and saturable absorption \cite{Boltasseva2011},
\cite{Alam2016,Jahani2016,west2010,Naik2013}.
Meanwhile, experimental ultrafast SPP photoexcitation on diamond
surfaces was not realized yet, even though their potential
imprinting in surface relief in the form of polarization-dependent
laser-induced periodical surface structures (LIPSS, surface
ripples) was numerously evidenced
\cite{Miyaji2008,Shinoda2009,Wu2003}. Such experimental studies
were devoted to the design and fabrication of bio-sensors,
employing the biocompatibility of the material, by ablative
surface nanostructuring of its surface with high-intensity
femtosecond (fs) laser pulses, assuring precise delivery of
energy, while precluding collateral thermal effects. In the case
of diamond, ultimate LIPSS periods of 100--125 nm on diamond-like
carbon for 800-nm fs-laser pulses \cite{Miyaji2008}, or even
50-100 nm on thin diamond films for 248-nm fs-laser pulses
\cite{Shinoda2009} (down to 30--40 nm on diamond-like carbon after
irradiation with 266-nm femtosecond pulses)
\cite{Wu2003,Yasumaru2003} were reported, empirically scaling as
the normalized laser wavelength $\lambda/2n$ ($n$ is the refractive
index of diamond), similarly to other dielectrics
\cite{Bonse2009,Gottmann2009}. However, despite some previous
attempts \cite{Miyaji2008,Derrien2013,Miyazaki2005}, the
underlying photoexcitation of diamond surface and SPP waves still
remain unexplained.

Generally, spatial LIPSS periods $\Lambda$ are known to depend on
the laser wavelength $\lambda$ and the polarization of the laser
electric field $\es$ and the number of laser pulses
\cite{Calvani2014,Sipe1983,Akhmanov1985_1,
Huang2009_1,Varlamova2006,Huang2009_2,Bonse2010,Tsibidis2015}. The
surface ripple period  can be slightly less than $\lambda $,
succeeding the in-plane weak interference of the incident
transverse fs-laser wave and almost transverse surface polaritons
\cite{Danilov2015}. These surface electromagnetic modes, residing
along the light cone line on dispersion curves for the metallic or
strongly photoexcited dielectric surface with its dielectric
permittivity $\varepsilon_m$ and its intact dielectric with its
dielectric permittivity $\varepsilon_d$ are photoexcited by the
fs-laser pump pulse via its scattering on permanent or
laser-induced (e.g., phase transition from diamond to glassy or
diamond-like carbon phase) cumulative surface relief roughness
\cite{Varlamova2006,Huang2009_2,Bonse2010,Tsibidis2015}, or prompt
laser-induced "optical roughness" \cite{Ionin2015}, if the
condition $\Re e[\varepsilon_m] \ll \Re e[\varepsilon_d$] is
fulfilled \cite{Huang2009_1,Ionin2013}. Meanwhile, in the
corresponding spectrally-narrow surface plasmon resonance,
occurring for the photoexcited surface at $\Re e[\varepsilon_m]=
-\Re e[\varepsilon_d$], the short-wavelength, longitudinal surface
plasmons can similarly interfere with the incident wave or among
themselves (for counter-propagating quasi-monochromatic surface
plasmons), inducing surface ripples with periods much lower than
$\lambda $ ( $\lambda $/2, $\lambda $/6, ..)
\cite{Huang2009_1,Borowiec2003,Golosov2011,Nathala2015}.
Importantly, in the former case, the surface polariton-mediated,
near-wavelength ripples are always oriented perpendicularly to
$\es$ (their wavevector $\kap || \es$), while the fine nanoripples
can be oriented in both ways, depending which -- red or blue --
shoulder of the surface plasmon resonance is involved
\cite{Kudryashov2015}. Laser exposure (the number of incident
pulses per spot, $N$) is known to influence LIPSS (both ripples
and nanoripples \cite{Nathala2015}) to much less extent, inducing
about 30{\%} reduction in their periods versus exposures,
increasing to $N \sim $ 10$^2$-10$^3$
\cite{Huang2009_2,Bonse2010,Tsibidis2015}. Other effects -- angle
of incidence/laser polarization \cite{Ionin2012}, intact
dielectric \cite{Golosov2009,Ionin2014,Bashir2015} indicate some
emerging possibilities in reduction of LIPSS periods, but should
be explored in details yet. Meanwhile, nanoscale hydrodynamics
instabilities of laser-induced surface melt were also considered
and explored as an alternative to the diverse electromagnetic
approaches \cite{Reif2011,Varlamova2013,Tsibidis2012}.

Since the prompt dielectric permittivity of the photoexcited
surface appears to be crucial for excitation either
near-wavelength surface polaritons, or sub-wavelength surface
plasmons, prompt photoexcitation (photoionization) of diamond,
directly affecting its dielectric permittivity, should be explored
in details. There are numerous semi-empirical approaches to
explain LIPSS formation e.g.
\cite{Sipe1983,vanDriel1982,Bonse2005,Jia2005,Dufft2009},
corroborating the experimental evidence, but no genuine
microscopic approach is invoked so far. The basic physical
processes involve excitation of electron-hole pairs, often
parameterized by Keldysh approximate formulas. Photoionization may
produce highly energetic electrons that collisionally ionize the
valence band and produce more electrons in the conduction band.
The multiplication of carriers may cause optical breakdown of bulk
diamond. The collective response of charge carriers screens out
the laser electric field inside the bulk when the number density
is sufficiently large. At some instant of time the bulk dielectric
function may become negative at the laser wavelength, allowing
excitation of SPP at the rough surface and LIPSS formation via the
optical interference mechanism. The dielectric properties of the
laser-irradiated material in most cases are parameterized with
Drude model
\cite{Huang2009_2,Golosov2011,Becker1988,Shimotsuma2003,
Bonse2009,Christensen2009}, which combines the ground state
response with the laser-induced free-carrier response. This model
usually requires three free parameters -- the number density of
electron-hole pairs, the free-carrier effective mass and the Drude
damping time, which are adjusted to fit experimental data. Ref.
\cite{Reitze1992,Sokol2000} proposed more elaborate model for the
optical dielectric function, which implements state- and
band-filling effects, renormalization of the band structure and
free-carrier response. The dielectric function of laser-excited
silicon was studied from first principles using the time-dependent
density functional theory (TDDFT) \cite{Sato2014}. A
distinguishing feature in the linear response of the photoexcited
silicon is a plasmon peak with large Drude damping time as short
as $\tau_e \sim 1$ fs, despite the neglect of collisional effects
in the TDDFT simulation. The real part of dielectric function was
well fitted by a Drude free-carrier response showing that $\Re
e[\varepsilon_m]$  is sensitive to the total number density of
excited electrons and not to the detailed distribution of
electron-hole pairs, while sensitivity to the nonequilibrium
distribution of the phototexcited carriers manifests in the
imaginary part of the dielectric constant. Subsequently, TDDFT was
applied to study ablation of silica subjected to ultrashort laser
pulses \cite{Sato2015}. The comparison between the estimated
surface ablation threshold and the experimental data suggests a
non-thermal mechanism in the laser ablation of silica by fs-laser
pulses, furthermore theoretical ablative crater depths agree with
the measured ones. The drawback of this approach is its limitation
to very short laser-matter interaction timescales (less than 10
fs).

In the present paper, we present theoretical and experimental
results for the laser ablation and LIPSS formation on diamond
surfaces subjected to normally incident 515-nm, 200-fs laser
pulses. Our theoretical modeling of LIPSS formation on diamond
surfaces is based on numerical solution of the time-dependent
Schr\"{o}dinger equation (TDSE) in bulk diamond subjected to a
single intense laser pulse. The theory describes the electron
dynamics quantum mechanically in the single-active-electron
approximation. Collisional de-excitation of the photoexcited
carriers and subsequent impact ionization are treated within rate
equation approach and an optical breakdown threshold is derived.
Due to the contribution of  the impact ionization the real part of the bulk
dielectric constant of the irradiated diamond becomes negative in
the trailing edge of the pulse resulting in plasma that is opaque
to the incident radiation. The inhomogeneous energy deposition in
the surface was modeled with the Sipe-Drude efficacy factor
theory \cite{Dufft2009,Bonse2009} in terms of time-dependent
dielectric function of free carriers. The applicability of this
efficacy factor theory for LIPSS formation in laser-irradiated
dielectrics was confirmed by numerical solutions of the Maxwell's
equations at statistically rough surfaces \cite{Skolski2012}. The
paper is organized as follows. In Sec. II we present the
theoretical approach to describe LIPSS formation on diamond
surfaces. Sec. III presents results for the ablative craters that
were experimentally produced on the surface of monocrystalline
diamond by multiple femtosecond laser pulses and the subsequent
emergence of fine and coarse surface ripples with the increasing
number of laser shots. The thresholds for surface ablation and
nano-structuring of diamond and their dependence on the
superimposed pulse number are obtained. The experimental data for
the observed surface ripple periods is consistently interpreted
within the Sipe theory based on free-carrier Drude response of the
laser-excited diamond. Sec. IV contains our main conclusions.

\section{Theoretical approach}

\subsection{Inhomogeneous energy deposition}

In order to model theoretically LIPSS formation in
femtosecond-laser-excited diamond, we apply the {\em ab initio}
theory developed by Sipe \cite{Sipe1983}.  In this picture, the laser beam is incident on a rough surface, the (permanent or laser-induced)
roughness is assumed to be confined within a surface region (selvedge) of thickness $l$ much smaller than the laser wavelength $\lambda$.
The optically-induced polarization in the selvedge generates surface-scattered waves that interfere with the
refracted laser beam leading to inhomogeneous energy deposition into the surface.
The inhomogeneous energy absorption can be described by the function
\begin{equation}
\label{eqA}
 A(\kap) \sim |b(\kap)| \eta (\kap;\kap_i),
\end{equation}
where $\kap_i$ is the component of the laser propagation wave vector parallel to the surface,
 $b(\kap)$ is a measure of surface roughness at
wave-vector $\kap=(\kappa_x,\kappa_y)$ and $\eta (\kap;\kap_i)$ is
an efficacy factor describing the contribution to the energy
absorption at the LIPSS wave vector $\kap$. The prediction of
Eq.\ref{eqA} is valid if the selvedge thickness is small
compared to the LIPSS period, i.e. $\kappa l \ll 1$ should be
satisfied. The efficacy factor essentially incorporates the
modification of the surface morphology and the variation of the
dielectric constant $\varepsilon$ of the photoexcited diamond. For
normally incident s-polarized laser pulse with $\kap_i={\bf 0}$,
the efficacy factor (as a function of the normalized wave-number
$\kappa=\lambda/\Lambda$) can be written as $\eta (\kap)=4\pi|\Re
e[\nu(\kap)]|$ with
\begin{equation}
\nu (\kap)=\left[ h_{ss} (\kappa)\left( {\frac{\kappa_{y}
}{\kappa}} \right)^{2}+h_{\kappa \kappa} (\kappa)\left(
{\frac{\kappa_{x} }{\kappa}} \right)^{2} \right] \gamma_t \vert
t_s \vert^{2},
\end{equation}
where the response functions $h_{ss}$ and $h_{\kappa \kappa} $
\begin{equation}
h_{ss} (\kappa)=2i \frac{ \kappa \kappa_m }{\kappa_v+\kappa_m},
\quad h_{\kappa \kappa}(\kappa)=2 \frac{\kappa_v
\kappa_m}{\varepsilon \kappa_v +\kappa_m},
\end{equation}
are given in terms of  the transient bulk dielectric  function
$\varepsilon(\omega;t)$ (cf. Sec. Optical properties),
$\kappa_v=\sqrt{\kappa^2-1}$ and
$\kappa_m=\sqrt{\kappa^2-\varepsilon}$, the Fresnel transmission
coefficient $t_s =2/(1+\sqrt{\varepsilon-1})$ in the absence of
the selvedge, the effective transverse susceptibility function
$\gamma_t =(\varepsilon-1)/4\pi
\left\{{\varepsilon-(1-f)(\varepsilon-1) [h(s)-Rh_{I} (s)]}
\right\}$,  the surface roughness characterized by shape $s$ and
filling $f$ factors, $R=(\varepsilon-1)/(\varepsilon +1)$ and the
shape functions $h(s)=\sqrt {s^{2}+1} -s,h_{I} (s)=(\sqrt
{s^{2}+4} +s)/2-\sqrt {s^{2}+1} $. When $\Re e[\varepsilon]<0$,
the response function $h_{ss}$ exhibits small kinks near the light
line $\kappa \approx 1 $, in contrast $h_{\kappa \kappa}$ exhibits
sharp resonance structure due to the excitation of surface
plasmons and diverges at the (complex) SPP wave-number
$\kappa_{SPP} =\sqrt{\varepsilon/(1+\varepsilon)}$.

\subsection{Photoexcitation}

Photoexcitation and the  dielectric response of laser-irradiated
diamond are treated in independent particle approximation based on
the 3D TDSE. In a long-wavelength approximation the light pulse is
represented by a spatially uniform time-dependent electric field and
velocity gauge is used throughout the calculations
\cite{Lagomarsino2016}. The static bulk band structure is
represented by the lowest 4 valence bands and 16 unoccupied
conduction bands. The Brillouin zone was sampled by a Monte Carlo
method using 2000 randomly generated $\ks$-points. The time step
for integration of the equations of motion was $\delta t=0.03$
a.u.

\begin{figure}
\includegraphics[width=.3\textwidth]{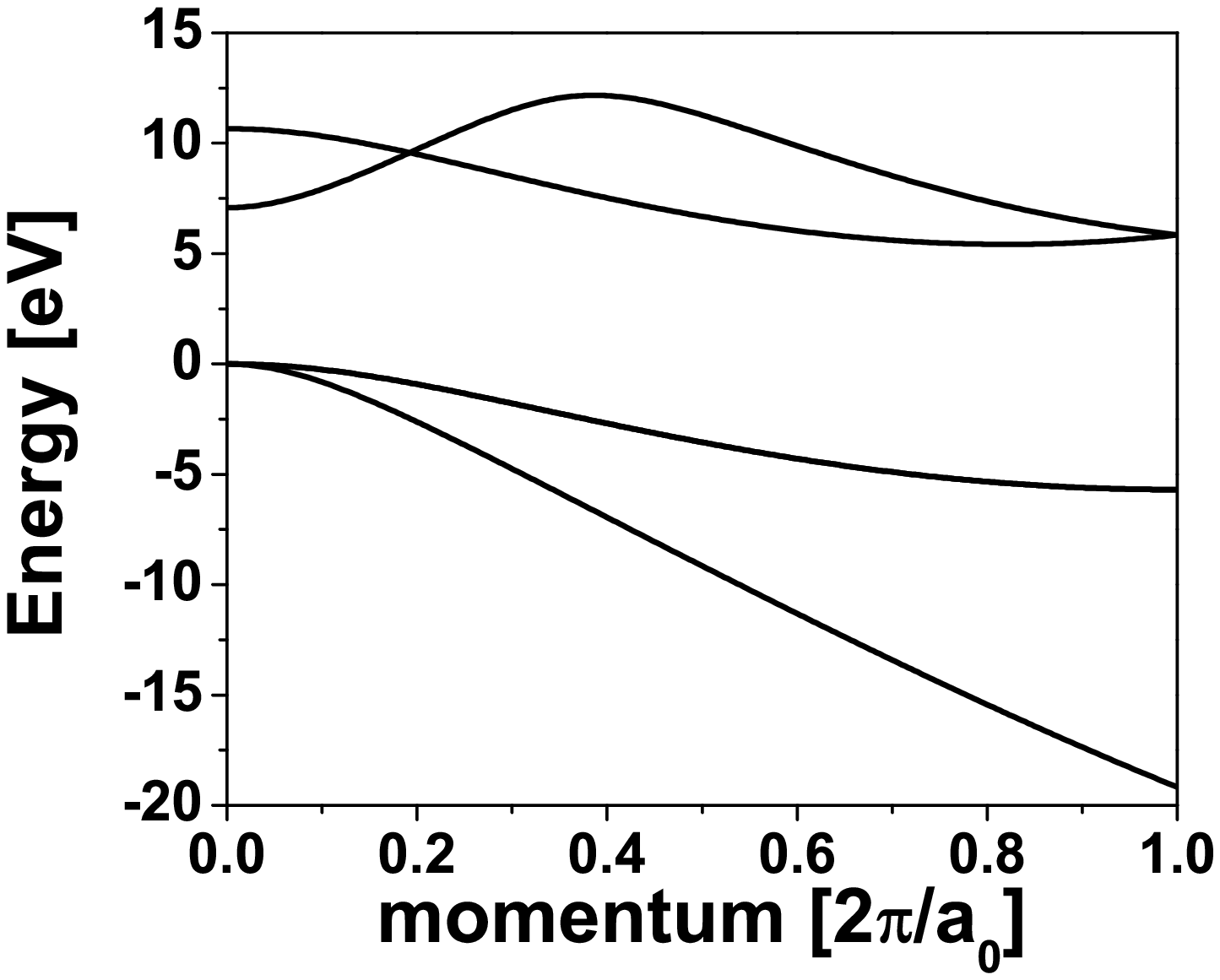}
\caption{Band structure of bulk diamond along the $\Delta $-line.
The momentum is measured in units of $2\pi /a_0$, where $a_0
=3.57\AA$ is the bulk lattice constant.} \label{F1}
\end{figure}

\begin{figure}
\includegraphics[width=.5\textwidth]{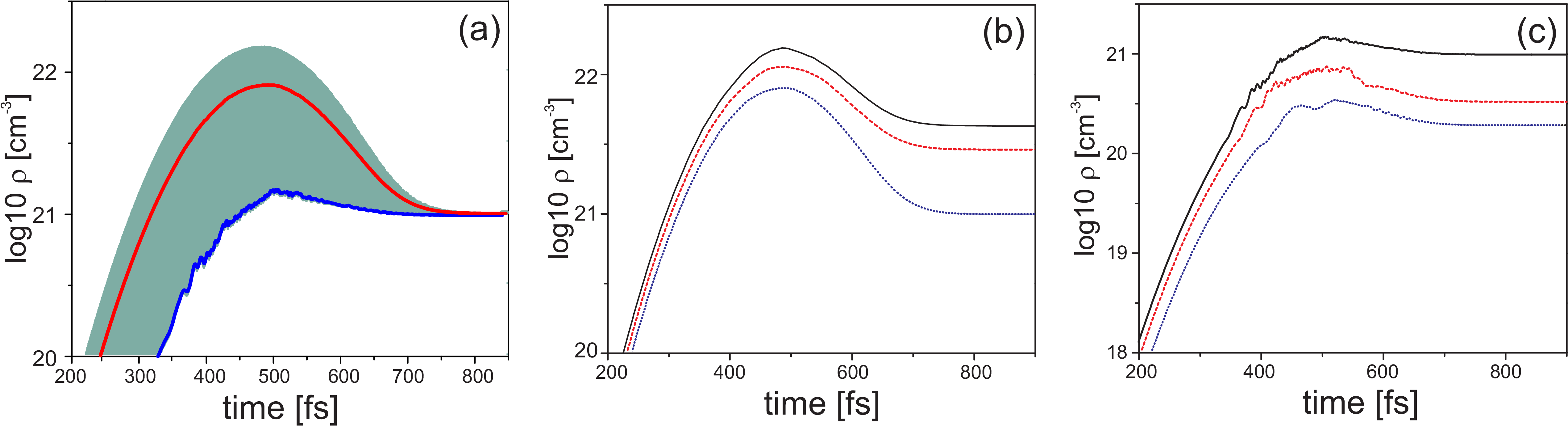}
\caption {Time evolution of the free-electron density in diamond
irradiated by 200fs laser pulse with intensity 30 TW/cm$^2$,
linearly polarized along the (1,1,1) direction. The red curve
shows the cycle-averaged electron density and the blue curve is
the electron density. (b) The cycle-averaged carrier densities for
intensity I=30, 40 and 50 TW/cm$^2$ are shown by the
dashed-dotted, dashed and solid lines, respectively. The position
of the pulse peak is indicated by the vertical dashed line and
Fig. (c) presents the number density of non-adiabatically excited
carriers for intensity I$=$30, 40 and 50 TW/cm$^2$} \label{F2}
\end{figure}

The static band structure along the $\Delta $-line is shown in
Fig.\ref{F1}. Carrier excitation occurs through the direct gap at
the $\Gamma$ point, however excitation into higher lying
conduction bands is also a relevant process for the considered
laser intensity range I$\in $[1,50] TW/cm$^2$.
During the irradiation of the diamond surface with pulsed 200
fs-laser, the total number of electrons generated into the
conduction band is given by a Brillouin zone integral
\begin{equation}
\label{eq1} \rho_e (t)=\sum_{\epsilon_n >0} \int_{{\rm BZ}}
\frac{d^3 \ks}{4\pi^3} f_{n \ks}(t) ,
\end{equation}
where $f_{n \ks}$ is the occupation number of the $n$-th
conduction band and $\ks$ is the crystal momentum. The electronic
excitation energy per unit cell is given by
\begin{equation}
\label{eq2} E_{ex} (t)=\sum_{\epsilon_n <0} \int_{{\rm BZ}}
\frac{d^3\ks}{4\pi^3}\left\langle \psi_{n\ks}(t)  | i
\partial_t | \psi_{n\ks} (t) \right\rangle -E_0,
\end{equation}
where  $\psi_{n\ks} (t)$ are the time-evolved Bloch orbitals of
valence electrons and $E_0$ is the ground-state energy. The time
evolution of the free-electron density is shown in Fig. \ref{F2}a,
for linearly polarized electric field along the (1,1,1) direction
with intensity 30 TW/cm$^2$. Carrier generation occurs
efficiently prior to the peak of the pulse. Transient charge
density oscillations following the laser period are due to quiver
motion of free electrons in the electric field. An electron-hole
plasma (EHP) with number density exceeding $10^{21}cm^{-3}$ is
established shortly after the peak intensity. The cycle-averaged
photoelectron yield, shown in Fig. \ref{F2}b, is a slowly varying
function of time. Carrier generation on the rising edge of the pulse competes
with recombination on the trailing edge of the pulse to determine
the final photoionization yield. Recombination of carriers becomes
unlikely with the increased laser intensity, cf. also Fig.
\ref{F2}b. The cycle averaged electron yield includes
contributions due to creation of real as well as virtual
electron-hole pairs. Since adiabatic evolution does not produce
any real excitation of the crystal, the carrier density should be
calculated with respect to adiabatically evolved ground state
orbitals that are obtained from the static Bloch orbitals with
shifted crystal momentum $\ks(t)=\ks+\ac(t)$, i.e. $\rho
(t)=\rho_e (t)-\rho_{ad} (t)$, where the adiabatic density is
\begin{equation}
\label{eq3} \rho_{ad} (t)=\sum_{\epsilon_n <0} \int_{{\rm BZ}}
\frac{d^3\ks}{4\pi^3} | \langle u_{n\ks} | u_{n\ks(t)} \rangle
|^{2},
\end{equation}
and $\{ |u_{n\ks} \rangle  \}$ are the
lattice-periodic Bloch states. The number density of photoexcited
carriers is shown in Fig. 2c. It can be seen that discarding
contributions of virtually excited electron-hole pairs leads to
reduction in the number density by an order magnitude near the
peak of the pulse.

\begin{figure}
\includegraphics[width=.4\textwidth]{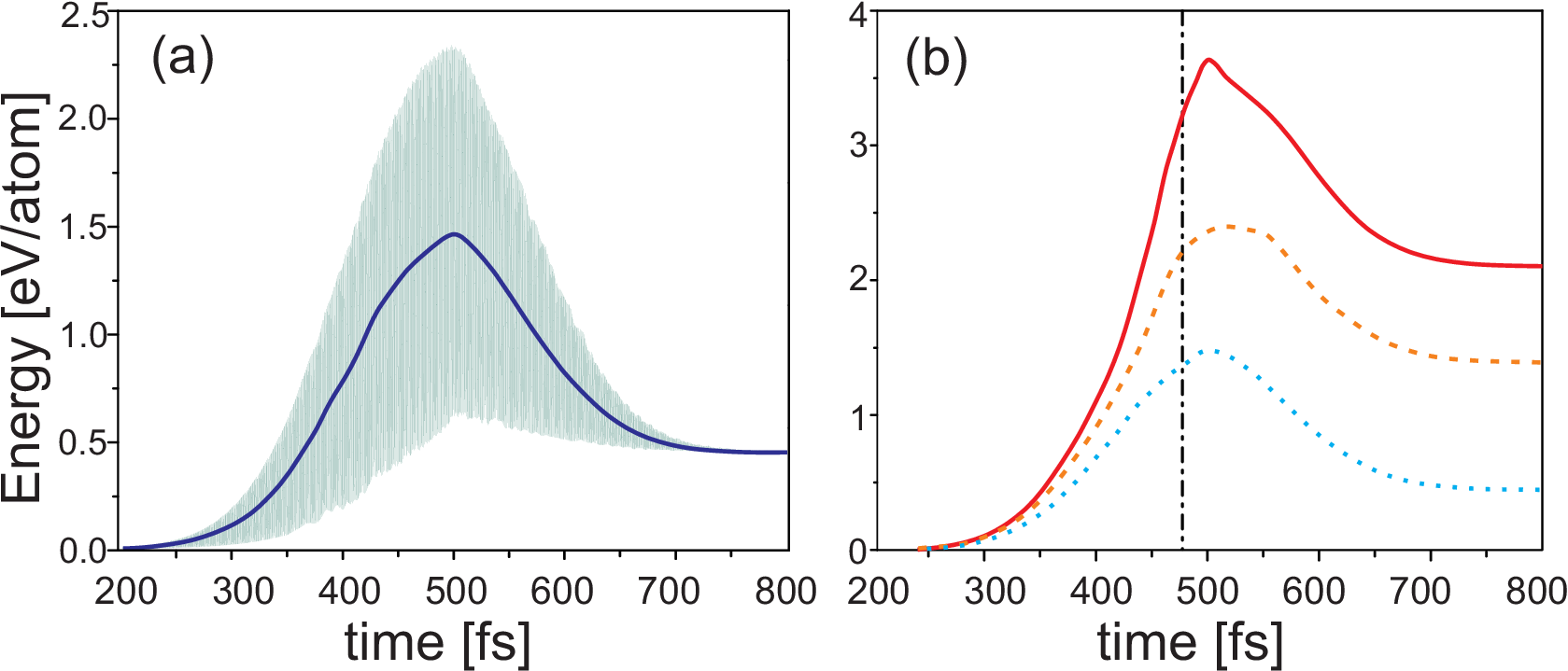}
\caption{ (a) Instantaneous excitation energy of electrons
interacting with 200fs laser pulse with intensity 30 TW/cm$^2$
(linearly polarized along the (1,1,1) direction), the green curve
shows the cycle-averaged electronic excitation energy. (b) The
cycle-averaged excitation energy for laser intensity I$=$30, 40
and 50 TW/cm$^2$ is shown by the dashed-dotted, dashed and solid
lines, respectively. The position of the pulse peak is indicated
by the vertical dashed line.} \label{F3}
\end{figure}

The electronic excitation energy is shown in Fig. \ref{F3}(a) for
laser intensity 30 TW/cm$^2$. The temporal variation of the
cycle-averaged energy gain follows closely the envelope of the
laser pulse during the first half of the driving pulse and
reaches 1.5 eV/atom at the peak of the pulse that is small as compared to the
cohesive energy of diamond 7.37 eV/atom. After the pulse peak,
electron-hole pairs recombine by transferring part of their energy
back to the radiation field. Energy exchange is not completely
reversible since the time delay in restoration of equilibrium
gives rise to a net energy gain of 0.5 eV per carbon atom after
the end of the pulse. The deposited energy increases steadily with
the increase of the intensity, i.e. for I$=$50 TW/cm$^2$, it
reaches 2 eV/atom. Since this excitation energy is still lower
than the diamond cohesive energy, Fig. \ref{F3}(b) shows that
ablation threshold is not reached up to I$=$50 TW/cm$^2$.

\subsection{Impact ionization and optical breakdown threshold}

\begin{figure}
\includegraphics[width=.5\textwidth]{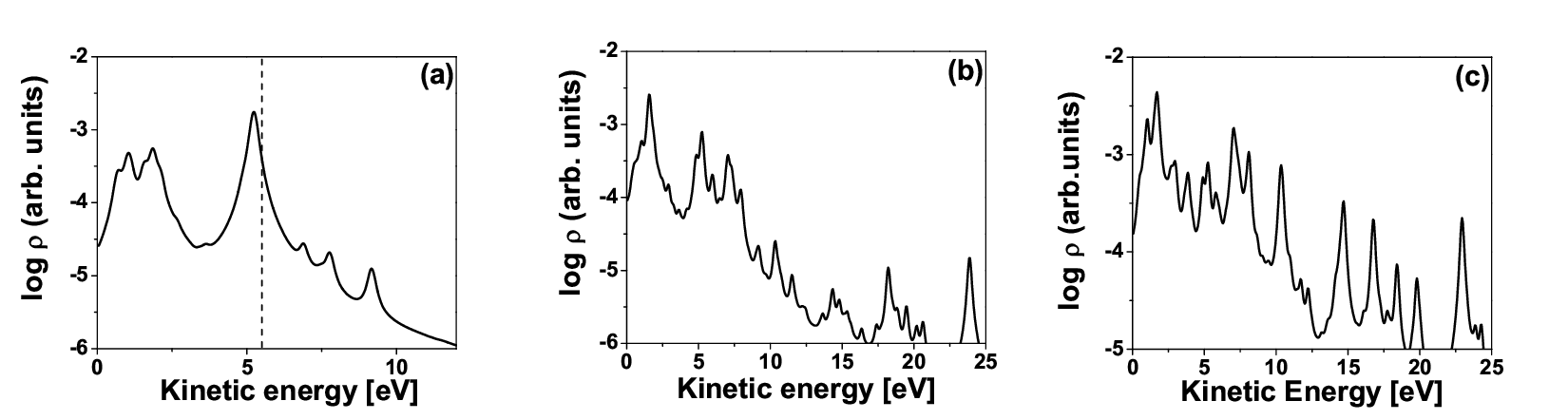}
\caption{Density of conduction states after the irradiation of
bulk diamond with 200fs laser pulse. The energy is measured
relative to the conduction band minimum. The laser intensity is
$I=$10, 20 and 30 TW/cm$^2$in Fig. (a-c), respectively. The
vertical dashed line in Fig.a indicates the threshold for impact
ionization} \label{F4}
\end{figure}

For the 200fs pulse duration and intensities lower than 50
TW/cm$^{2\, }$ the electron density produced by photoionization is
below the critical one.  That suggests that impact ionization is the
relevant process that determines the optical breakdown threshold.
In Fig.\ref{F4} (a-c) we plot the density of conduction states
after the end of the pulse. It is seen that the laser has created electron-hole
pairs with well-defined energies. This non-thermal distribution relaxes towards the equilibrium
Fermi-Dirac distribution on a time scale ranging over few tens of a femtoseconds
to a picosecond \cite{Yoffa1980,Fann1992} without changing the electron number density.
Photoelectrons are excited into the lowest  conduction band across the direct gap
(with energies 2 eV above the conduction band
minimum) and substantial fraction of carriers occupy higher lying conduction bands with
energy above threshold for impact ionization (specified by the indirect gap 5.4 eV). These
highly energetic electrons may collisionally de-excite to lower energy states
and their excess energy is spent to promote valence electrons into the conduction band.

We further assume that the time evolution of the electron density
is governed by a rate equation  \cite{Apostolova2000,Stewart1995}
\begin{equation}
\label{eq4} \frac{d\rho }{dt}=G(t)-R(t)+w_{imp} (I) \rho
\end{equation}
including carrier generation $G(t)$ and recombination $R(t)$ rates
supplemented by an intensity-dependent impact ionization rate
$w_{imp} $ obtained as a weighted-average of the field-free
ionization rate
\begin{equation}
\label{eq5} w_{imp}(I)=\frac {\int_{\epsilon_i }^{\infty}
d\epsilon \rho (\epsilon;I) P_{imp} (\epsilon) } {\int_0^{\infty}
d\epsilon \rho (\epsilon;I) }
\end{equation}
here $\rho (\epsilon;I)$ is the density of conduction states after
the end of the pulse (cf. Fig.\ref{F4}a-c),
\begin{equation}
\rho (\epsilon;I)=\sum_{\epsilon_n >0} \int_{{\rm BZ}} \frac{d^3
\ks}{4\pi^3} f_{n \ks} (I) \delta (\epsilon -\epsilon_{n \ks}),
\end{equation}
$P_{imp} (\epsilon)=P_0 (\epsilon -\epsilon_i)^{4.5}$ is the
energy-dependent impact ionization rate for diamond, $\epsilon_i$
is the threshold for impact ionization (5.42 eV) and
$P_0=3.8\times 10^{10} s^{-1}eV^{-4.5}$ \cite{Watanabe2004}.

\begin{figure}
\includegraphics[width=.5\textwidth]{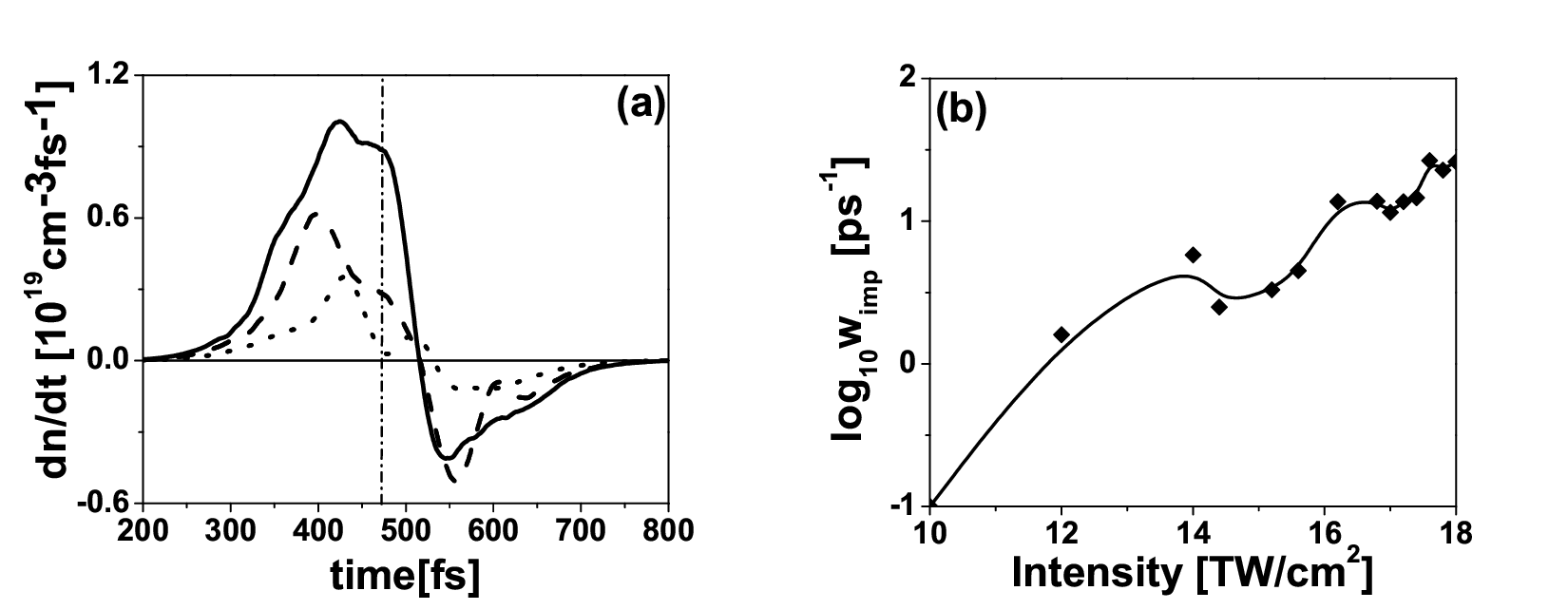}
\caption{Fig. (a) Time-dependent rates including carrier
generation (positive part) and laser-induced recombination (negative part). The laser
intensity is 10 TW/cm$^2$(dotted), 20 TW/cm$^2$ (dashed)
and 30 TW/cm$^2$ (solid line). Fig.(b) shows the
intensity-dependent impact ionization rate.} \label{F5}
\end{figure}

In contrast to the standard perturbative result based on Keldysh
theory valid for monochromatic laser radiation the calculated  carrier generation
and recombination rates shown in Fig.\ref{F5}a do not follow the
temporal profile of the laser pulse. This result suggests that the
pulse shape and pulse duration are relevant control parameters for
non-adiabatic electron dynamics in the laser irradiated diamond.
The key features  are generation of dense plasma 50fs
prior to the pulse peak and subsequent laser-induced recombination
of electron-hole pairs in the trailing edge of the pulse.

\begin{figure}
\includegraphics[width=.3\textwidth]{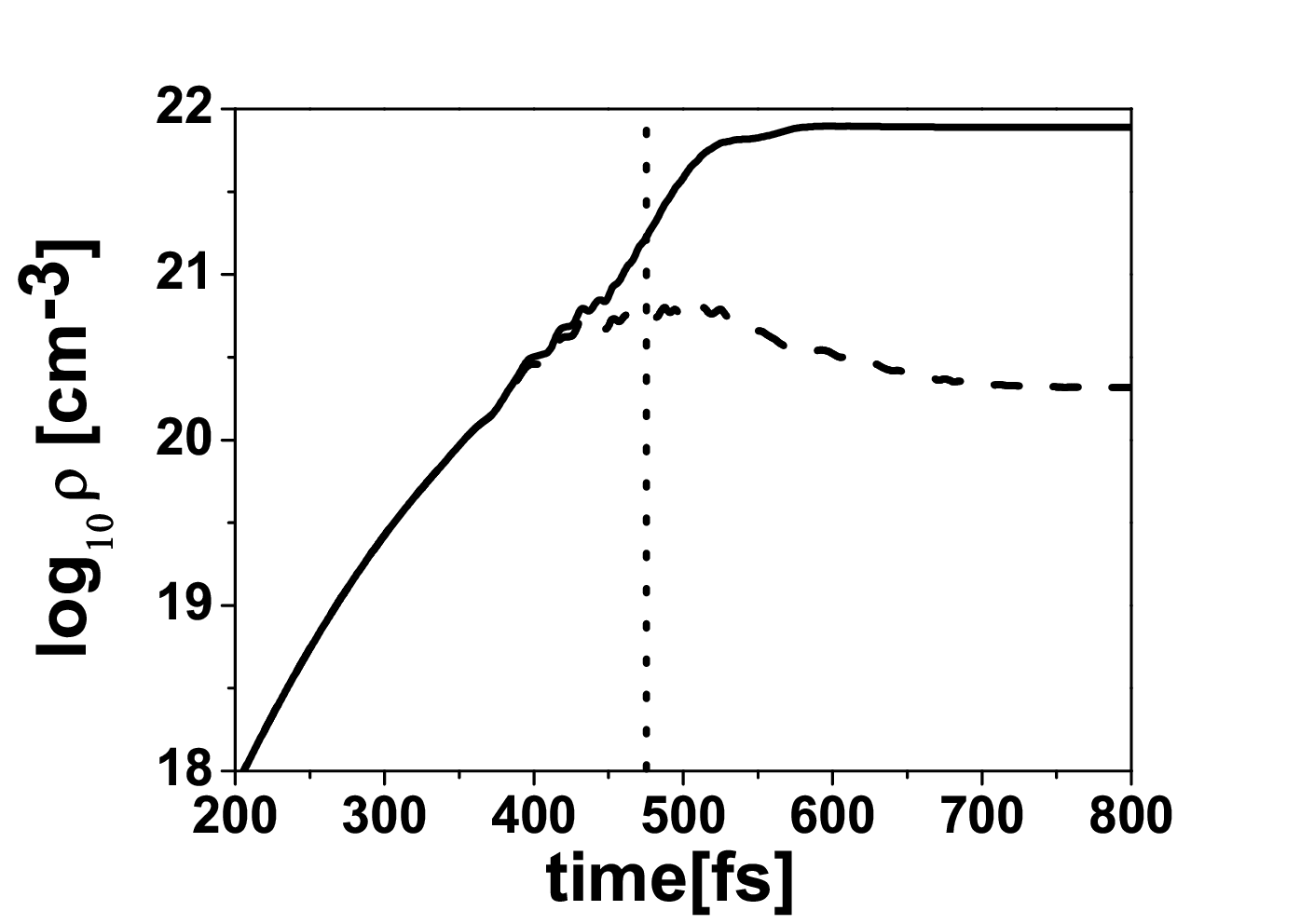}
\caption{Conduction electron density due to photoionization only
(dashed line) and including the impact ionization (solid line).
The laser intensity is 18 TW/cm$^2$. The vertical dotted line
indicates the position of the pulse peak.} \label{F6}
\end{figure}

Fig.\ref{F5}b shows the impact ionization rate that depends in
highly non-linear way on the laser intensity. This non-linear
and non-monotonic intensity-dependence reflects the population
of higher-lying conduction bands (cf. also Fig.\ref{F4}).
It is seen that the impact ionization rate reaches few tens of inverse picosecond for
$I > 15$ TW/cm$^2$. In Fig.\ref{F6} we plot the EHP density with and
without the impact ionization term. This comparison demonstrates that
photoionization produces the seed electrons needed for the impact
ionization on the rising edge of the pulse and then the conduction electron
density grows exponentially after the pulse peak resulting in
dense plasma (with density 10$^{22}$ cm$^{-3})$ 50fs after the
peak of the pulse.

\subsection{Optical properties}

Since the absorption of the femtosecond laser pulses in
diamond results in the generation of nearly free electrons in the
conduction band on timescales smaller than the electron-phonon
relaxation time \cite{groenveld1995}, we describe the linear response of the photoexcited
diamond by a free-carrier Drude response \cite{Bonse2009}
using of time-dependent plasmon-pole-approximation for the
density-density correlation function of the Coulombically
interacting electron gas \cite{Sayed1995}
\begin{equation}
\label{eq6} S(t,t')=-\theta (t-t')\omega_p^{3/2}
(t')\omega_p^{-1/2} (t)\sin \int_{t'}^t d\tau \omega_p
(\tau) ,
\end{equation}
where $\theta(t)$ is the Heaviside step function, $\omega_p
(t)=\left( {\frac{\rho(t)}{\varepsilon_0 m_e}} \right)^{1/2}$ is
the bulk plasma frequency, $\varepsilon_0$ and $m_e $ are the
vacuum permittivity constant and the free-electron mass,
respectively. In long wavelength approximation the spatial
dispersion of the bulk plasmon is neglected. The Fourier
transformation of the correlation function is the transient
frequency dependent inverse dielectric function of the
free-electron plasma
\begin{equation}
\label{eq7} \varepsilon^{-1}(\omega;t)=1+\int\limits_{-\infty }^t
{dt'e^{i(\omega +i\delta )(t-t')}S(t,t')} ,
\end{equation}
where $\delta=1/\tau_e$ is a free-carrier polarization dephasing
rate, which we shall treat as a free parameter. If the time delay
in the build up of screening in the optically excited plasma can
be neglected, the classical Drude dielectric function is recovered
$\varepsilon^{-1}(\omega;t)=\omega^2/(\omega^2-\omega^2_p(t))$
with parametric time dependence of the bulk plasma frequency.
\begin{figure}
\includegraphics[width=.5\textwidth]{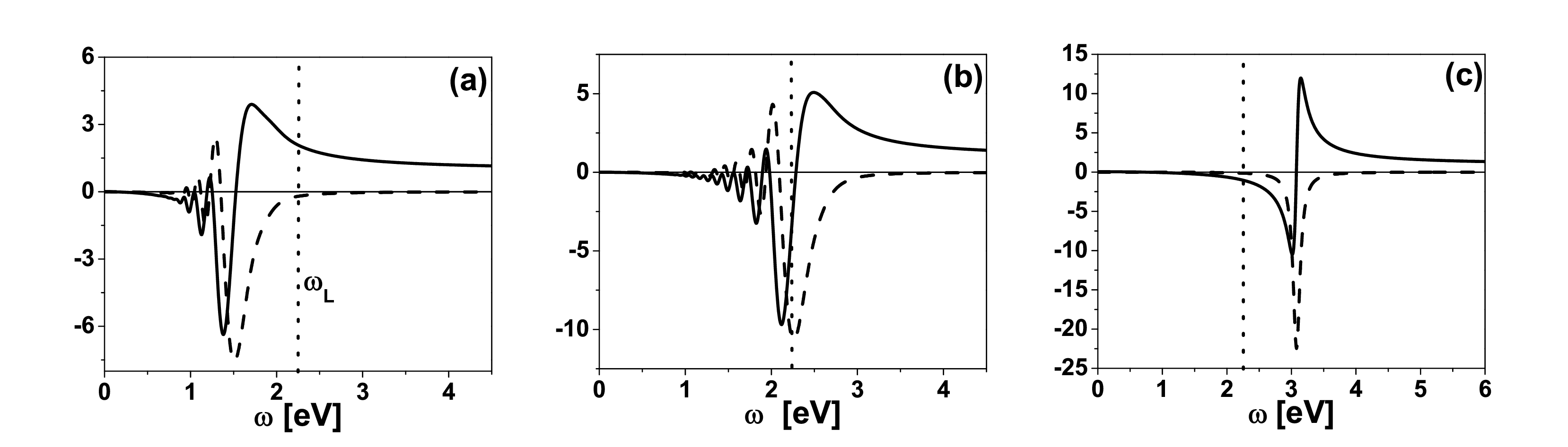}
\caption{Frequency dependence of the real (solid line) and
imaginary part (dashed line) of the inverse dielectric function of
photoexcited carriers subjected to 200 fs laser pulse with
intensity 18 TW/cm$^2$. The time interval is measured relative to
the peak of the pulse ($t=0$). In Fig. (b) $t = 50$ fs, and in
Fig. (c) $t=250$ fs. The photon energy is indicated by the
vertical dotted line.} \label{F7}
\end{figure}

In Fig.\ref{F7}a-c, we plot the real and imaginary parts of the
dielectric function for laser intensity 18 TW/cm$^2$. The
screening charge density accumulates during the first half of the
pulse ($t < 0$). Over that time interval the laser frequency is
above the plasma frequency and the diamond surface remains
transparent to the incident radiation. The frequency dependent
dielectric function displays oscillations in the spectral range
below the laser frequency due to the time lag in the
build up of screening. Because of the impact ionization, the laser
frequency falls off below $\omega_p$ after the pulse peak ($t= 25$
fs) when the plasma is reflective for the incident radiation and
an optical breakdown threshold is reached. In this regime, the
dielectric function essentially exhibits the Drude form with
time-dependent bulk plasma frequency $\omega_p(t)$.  For the
transiently increasing carrier density, $\Re e[\varepsilon]$
passes the narrow surface plasmon resonance at
$\Re e[\varepsilon]=-1$, with $\kappa_{SPP}  \gg  1$ and
$\Lambda/\lambda \ll 1$, and becomes large and negative in the
trailing edge of the pulse ($t> 100$ fs) with
$\Im e[\varepsilon] > 0$, with corresponding $\kappa_{SPP}  \ge
1$. During this plasmonically-active phase of the laser-irradiated
diamond the SPP-laser interference mechanism of inhomogeneous
energy deposition is effective and leaves permanent imprints on
the surface morphology after the conclusion of the pulse.

\section{Comparison of theory and experiment}

\subsection{SPP-mediated surface ripples in diamond: generation and
characterization}

SPP-mediated surface ripples were produced on a 0.5-mm thick plate
of monocrystalline A-type diamond nanostructured with the help of
laser nano/microfabrication workstation  \cite{Danilov2016}. The
sample was arranged on a three-dimensional motorized translation
micro-stage under PC control and moved from spot to spot to make
possible ablation of its fresh spots at variable number of pulses
N. Single- and multi-shot ablation of the sample was produced by
515-nm, 220-fs TEM$_{00}$-mode laser pulses weakly (NA $\approx $
0.1) focused into a focal spot with a 1/e-radius about 5.5 $\mu $m
at the energy $E=$ 3.4 $\mu $J (the peak intensity $I_0 \approx $ 10 TW/cm$^{2})$. The resulting single- and
multi-shot craters were characterized by means of a scanning
electron microscope (JSM JEOL 7001F)) and a Raman microscope
U-1000 (Jobin Yvon) at the 488-nm pump laser wavelength.

\begin{figure}
\includegraphics[width=.5\textwidth]{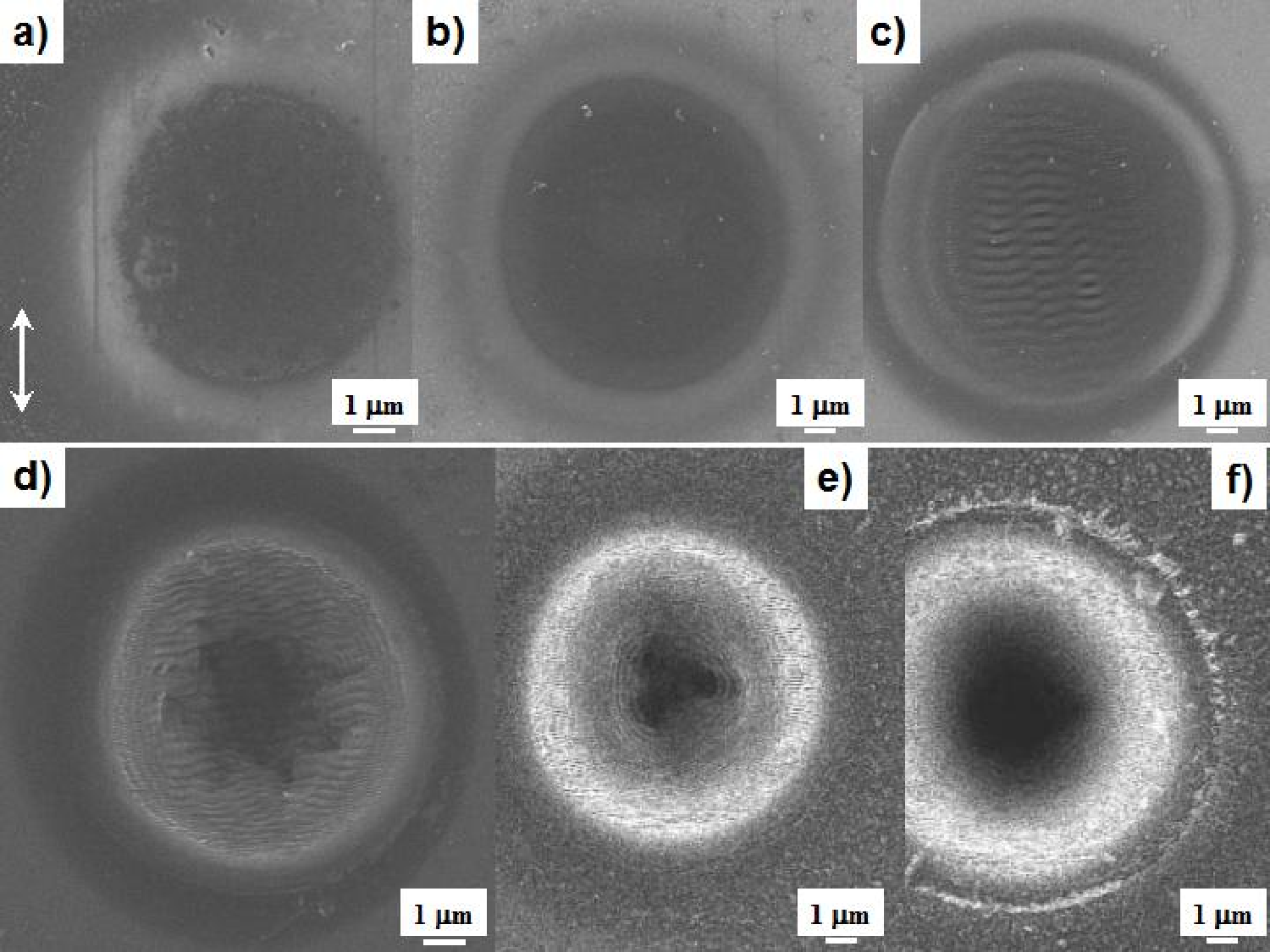}
\caption{SEM images of ablative craters on the diamond surface for
N $=$ 1 (a), 2 (b), 30 (c), 100 (d), 300 (e) and 1000 (f) pulses.
The scale bars are somewhat different on each image.} \label{F10}
\end{figure}

\begin{figure}
\includegraphics[width=.5\textwidth]{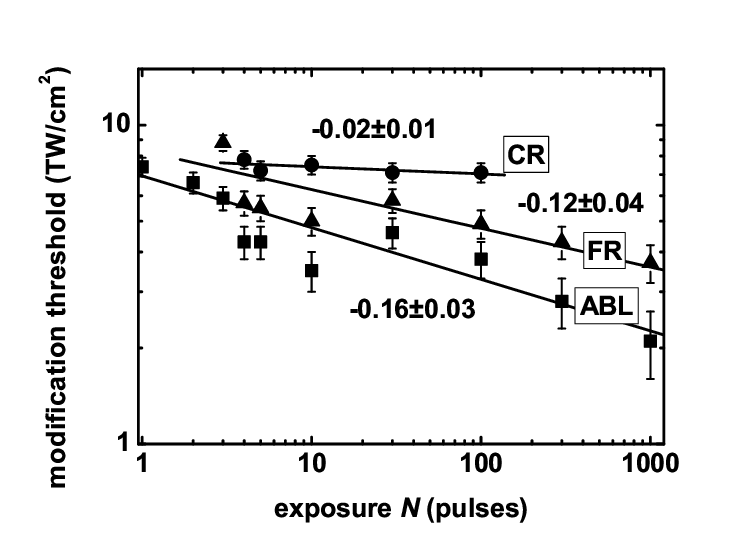}
\caption{N-dependent variation of ablation (ABL) and
nanostructuring (coarse and fine ripples, CR and FR) thresholds
with the corresponding linear fitting lines and slopes.}
\label{F11}
\end{figure}

Surface ablation of the crystalline diamond occur for fs-laser
intensities, exceeding the single-shot ablation threshold
$I_{abl}$(\ref{eq1}) $\approx $ 14.4 TW/cm$^2$ (Fig.\ref{F10}),
but for longer exposures $N \gg 1$ the
threshold intensity decreases down to $I_{abl}$(1000) $\approx $
2.1 TW/cm$^2$, following the well-known accumulation
relationship $I(N)= I$(\ref{eq1})$N^{-\alpha}$, where
$\alpha =  - 0.16\pm 0.03 $(Fig.\ref{F11}). The spallative origin
of the external crater is clearly seen as the sharp crater edge in
Fig.\ref{F10}f, however, at higher exposures another ablation
mechanism -- apparently, phase explosion -- comes into play for
$I_0 \gg I_{abl}(N)$, forming the deep central dips and destroying
the intermediate LIPSS (Figs.\ref{F10}d-f).

\begin{figure}
\includegraphics[width=.5\textwidth]{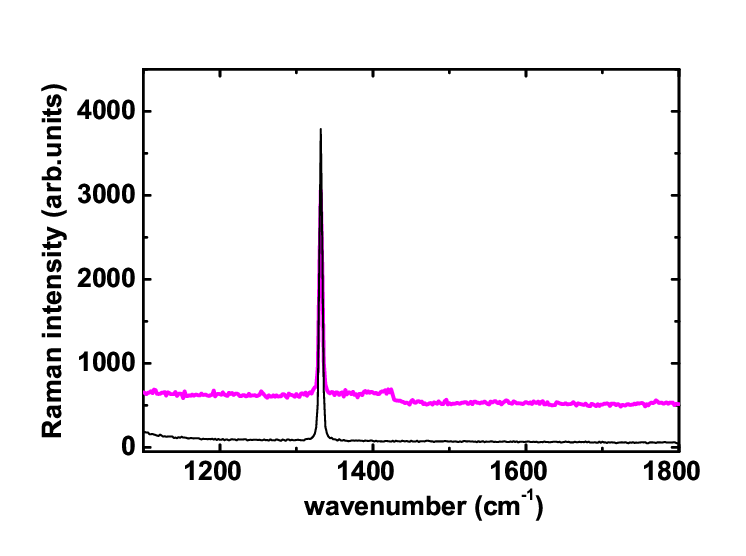}
\caption{Raman spectra of the D-band for the reference diamond
spot (bottom black curve) and the 10-shot crater (top purple
curve).} \label{F12}
\end{figure}

The observed cumulative decrease of the surface ablation threshold
can be related, e.g., to the increasing coloration shown by SEM as
darker ablated spots in Fig.\ref{F10}, as well as to stress,
structural damage and ablative modification of the crater surface
(Fig.\ref{F10}a-c). In particular, micro-Raman characterization of
the craters, exhibiting only slightly displaced D-band with
low-intensity background (Fig.\ref{F12}), is in agreement with
some previous fs-laser nanostructuring studies on diamond surfaces
\cite{Shinoda2009,Calvani2014}, showing rather clean
nanostructured surfaces. The low-intensity ultrabroad (1100-11400
cm$^{-1})$ difference spectral band is known to yield from
luminescence of nanoscale clusters  \cite{Tan2013},rather than
from the pump radiation, since both these spectra exhibit similar
D-band intensities and the pump radiation was cut in the
experiments in the same way. Moreover, the displaced ($\Delta \approx $ 0.1 cm$^{-1})$ D-band shown by the corresponding
bipolar band in the difference Raman spectrum (Fig.\ref{F12}),
indicates the internal residual stresses $\sim 0.3$ kbar,
according to the known calibration coefficient for this band
$\approx $ 0.336 cm$^{-1}$ / kbar \cite{Mitra1969}.

\begin{figure}
\includegraphics[width=.5\textwidth]{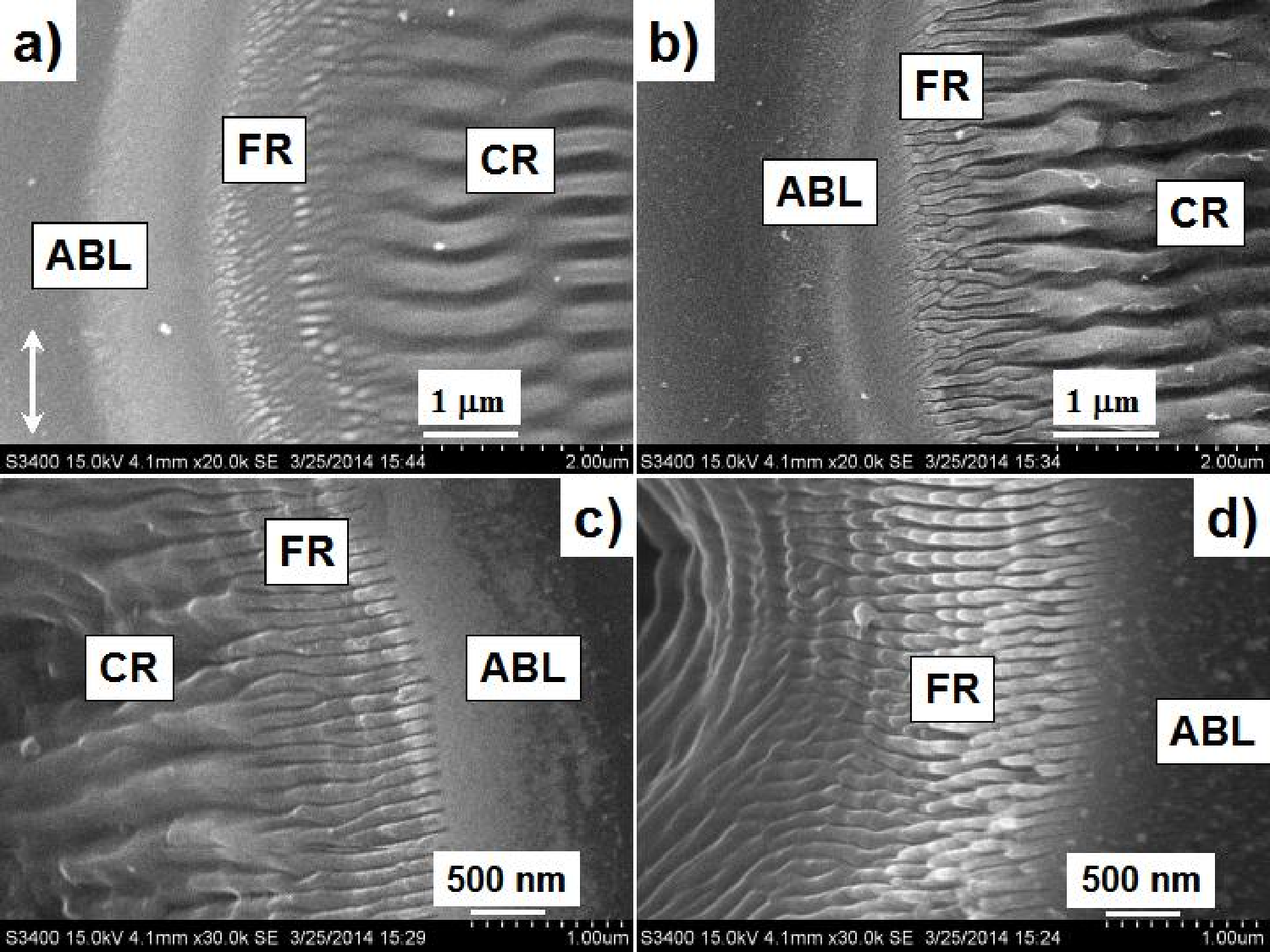}
\caption{SEM images of ablation crater edge (ABL), fine (FR) and
coarse (CR) rippled regions within the craters on the diamond
surface for N $=$ 10 (a), 30 (b), 100 (c), and 300 (d) pulses. The
scale bars are different on each image and the bi-lateral arrow in
a) shows the laser polarization.}  \label{F13}
\end{figure}

Fine and coarse surface ripples appear within the ablative
craters, starting from $N > 3$, inside the surface
regions limited by $I_{FR}(N)  \le I \le I_{CR}(N)$
and $I_{CR}(N) \le I \le I_0$ (Fig.\ref{F13}),
respectively. These thresholds exhibit two different trends with
the increasing exposure $N$ -- monotonous decrease for $I_{FR}(N)$
scaling as $\beta = -0.12 \pm 0.04$ for intensities from $8.8\pm
0.5$ to $3.14 \pm 0.5$ TW/cm$^{2}$ (Fig.9) and almost no variation
for $I_{CR}(N)$  (scaling as $\gamma = -0.02\pm 0.01$) for
intensities in the range from $14.8\pm 0.5$ to $14.1\pm 0.5$
TW/cm$^{2}$ (Fig.\ref{F11}). The minor variation of $I_{CR}(N)$
potentially indicates that the CR are formed due to scattering
mechanism, i.e. independent on the surface absorption, while
surface absorption is more crucial for the formation of fine
ripples.

Moreover, in comparison to fine ripples with threshold $I_{FR}(N) \ge $ $I_{abl}(N)$, coarse ripples, having considerably
higher threshold $I_{CR}(N) \gg I_{abl}(N)$, disappear in the central crater part for $N > 100$ because of the pronounced ablation in this region
(cf. Fig.\ref{F10} and Fig.\ref{F13}). Fig. \ref{F13} shows that
considerable CR erosion is present for $N=$ 30 and 100.

\begin{figure}
\includegraphics[width=.5\textwidth]{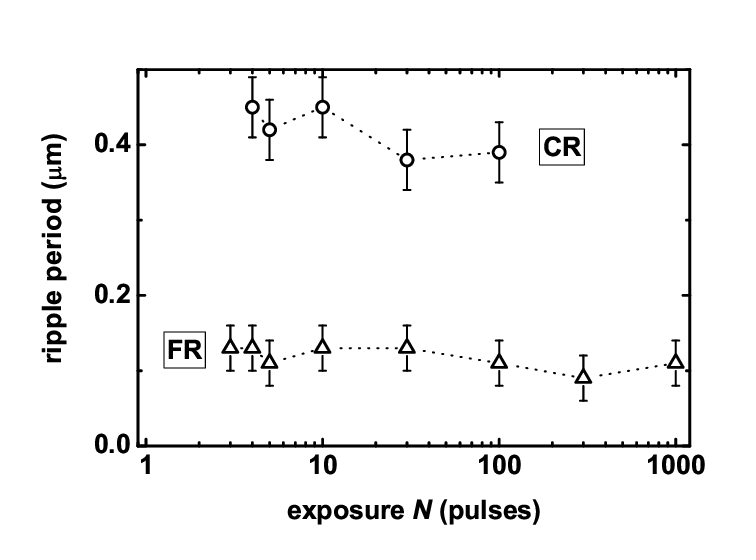}
\caption{ $N$-dependent variation of CR and FR periods.}
\label{F14}
\end{figure}

Most importantly, the small difference between the CR periods
($\le $0.45 $\mu m$, wavenumber $\ge  2.2 \mu$m$^{-1})$ and the
laser wavelength $\lambda$ (0.515 $\mu$m, wavenumber
$\approx $ 1.9 $\mu$m$^{-1})$ points out that long-wavelength
micron-scale ($\sim 3 \mu$m) perturbations of surface relief
(permanent or cumulative ones -- e.g., the spallative crater edge
for $N \ge $1) or optical characteristics (prompt or
cumulative ones) \cite{Ionin2015,Ionin2013} are responsible for
excitation of the underlying near-wavelength plasmon-polaritons.
The corresponding FR and CR periods decrease versus $N$ -- from
$0.13 \pm 0.03$ till $0.09 \pm 0.03 \mu$m and from $0.45\pm 0.04$
till $0.38\pm 0.04 \mu$m (Fig.\ref{F14}), respectively, in
agreement with cumulative trends known for FR and CR
\cite{Huang2009_2,Bonse2010,Tsibidis2015,Ionin2015}

\subsection{Interpretation of LIPSS as imprints of transient SPP modes}
\begin{figure}
\includegraphics[width=.5\textwidth]{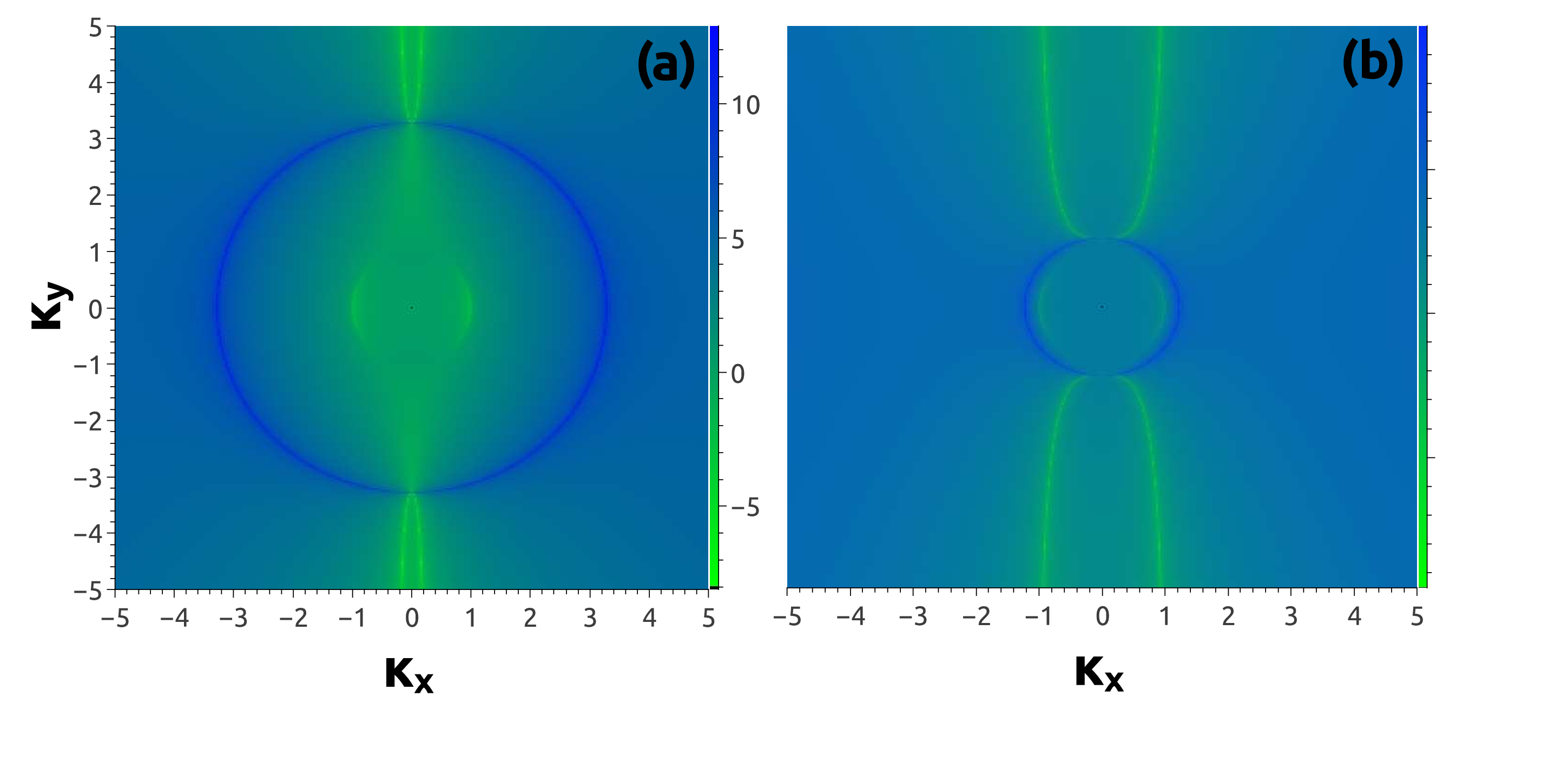}
\caption{2D intensity map of the logarithm of the transient
efficacy factor of laser-irradiated diamond, as a function of the
normalized (to the laser wavelength) LIPSS wave vector components
$(\kappa_x,\kappa_y)$, for Instantaneous  bulk plasma frequency
(a) $\omega_p(t)=1.5 \omega_L$ and (b) $\omega_p(t)=1.43
\omega_L$. The laser beam is linearly polarized along the y-axis
and is normally incident to the surface.} \label{F8}
\end{figure}

To make possible identification and interpretation of
experimentally obtained SPP modes  we plot the
efficacy factor as a function of the wave vector $\kap$ in a
narrow laser intensity range above the optical breakdown
threshold in Fig.\ref{F8} a-b. The transient bulk dielectric function was evaluated at
the laser wavelength, i.e.
$\varepsilon(t)=\varepsilon(\omega_L;t)$. The surface roughness
was modeled as a collection of spherically-shaped islands
corresponding to standard values $s=0.4$ and $f=0.1$ for the shape
and filling factors respectively. For normally incident light
pulse, the numerical results are weakly dependent on the specific
parameters describing surface morphology and therefore the transient
dielectric constant is the most significant in determining the efficacy
factor. Here we demonstrate that the main features in the
inhomogeneous energy deposition in the surface as represented by
the position of the peaks of the transient efficacy factor are in
correspondence with the experimentally observed LIPSS periods.

\begin{figure}
\includegraphics[width=.5\textwidth]{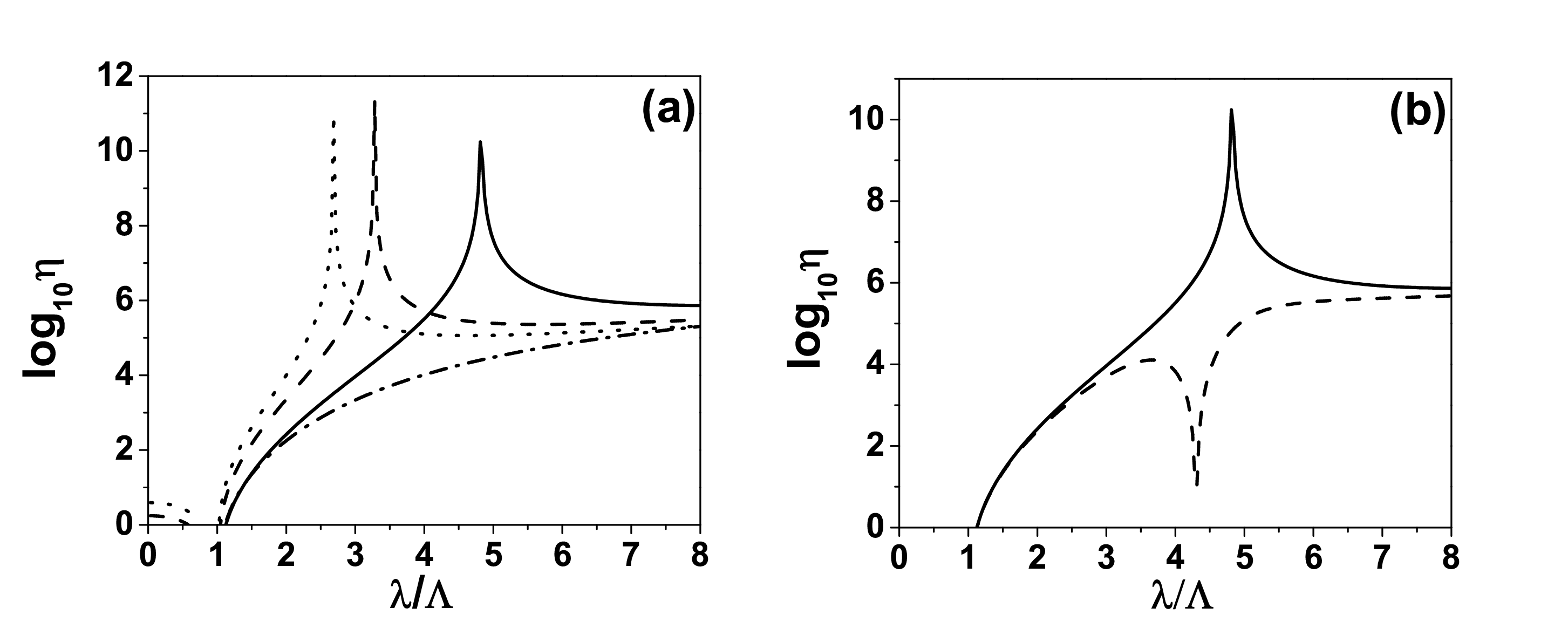}
\caption{Transient variation of the efficacy factor in the
direction perpendicular to the laser polarization. In Fig. (a),
the dashed-dotted line the EHP frequency $\omega_p(t) =1.4
\omega_L$ is just below the SPR excitation threshold, the solid
line represents resonant excitation of surface plasmons
corresponding to instantaneous bulk plasma frequency $\omega_p(t)
=1.43 \omega_L$, the transient increase of the free-carrier
density with $\omega_p(t)=1.45 \omega_L$ (dashed line) and
$\omega_p(t)=1.47\omega_L$ (dotted line) results in red-shift of
the surface plasmon peak towards the light line and formation of
near wavelength ripples. Fig.(b) demonstrates the dependence of
the efficacy factor on the Drude damping time $\tau_e =100$ fs
(solid line) and $\tau_e=10$ fs (dashed line). The laser beam is
linearly polarized along the y-axis and is normally incident to
the surface, $\lambda$ and $\Lambda$ designate the laser
wavelength and the LIPSS period, respectively.} \label{F9}
\end{figure}

In a very narrow laser intensity range, when the laser frequency
nearly matches the surface plasma frequency (Fig. \ref{F8}a), the
efficacy factor has large contribution due to excitation of the
surface plasmon resonance (SPR). In this early stage, the
spatial extension of the electromagnetic field inside the bulk
associated with SPP is determined by the SPR decay constant
$\kappa_m$, which at short wavelengths $\kappa_m \rightarrow 1/l$
defines a skin-depth $l_s = 1/\kappa_m \sim l$ leading to strong
concentration of the electromagnetic field in the thin selvedge
region. SPPs need finite time to build up to incorporate the details of
the surface relief and interfere with the laser to modify the
Fourier components of the surface roughness function $b(\kap)$ via
periodic laser ablation. In this regime the deposition of laser
energy into the surface plasmon wavevector causes formation of
fine ripples with spatial periods around 100 nm, as observed in
the periphery of the ablative craters, cf.  Fig.\ref{F13}.
The transverse-magnetic characteristic of the SPP determines the orientation
of the surface ripples. At a later time, the transiently increasing number of conduction
electrons makes the dielectric constant large and negative at the
laser wavelength, the intensity map of $\eta(\kap;{\bf 0})$
shrinks and concentrates on the outer part of the circle
$\kappa=1$ (cf. Fig \ref{F8}b) which clearly can be associated
with the formation of near-wavelength surface ripples oriented
perpendicularly to the laser polarization. At the longer
wavelengths with $\kappa \rightarrow 0$ the skin depth
$l_s=c/\omega_p \approx 80$ nm  is much smaller than the laser
wavelength. Therefore, above the SPR excitation threshold, the
transiently increasing carrier density results in a shift of the
SPP wave number from the high spatial frequency region towards the
light line (also causing expansion of the skin depth), and this
red shift is highly sensitive on the carrier density (or laser
intensity), cf. Fig.\ref{F9}a. The surface plasmon peak in the
efficacy factor is also affected by the relaxation time parameter
$\tau_e$ as shown in Fig. \ref{F9}b. If $\tau_e$ is decreased to 10
fs, the surface plasmon cusp turns into a dip, which hinders the
efficient energy absorption at the surface plasmon wavevector.
This dependence suggests that the Drude carrier relaxation time
parameter influences prompt feedback mechanisms involved in the
formation of surface ripples. Indeed in the high-frequency limit
with $\omega_L \tau_e \gg 1$, the metalized surface behaves as a nearly
ideal inductor, while in the low-frequency limit $\omega_L \tau_e \ll 1$,
the resistive Ohmic losses result in electron heating in the skin layer.

Because the efficacy factor theory does not fully account for interpulse
feedback processes that are undoubtedly important in the detailed
development of morphological features on the diamond surface our
theoretical results are not directly applicable to the multipulse
phase of LIPSS formation. However the experimental data shows that
once surface ripples are formed, exposure by subsequent pulses has
little effect on their spatial period and location, thus LIPSS
formation should be possible already for a single-pulse
irradiation, provided that SPP can be excited, e.g., by surface
defects \cite{Jia2015}. Once LIPSS are formed, the spectrum of the
surface roughness, $b(\kap)$ contains peaks at the SPP
wavenumber causing enhanced inhomogeneous energy deposition and
further growth of LIPSS as is also evidenced from the SEM images in
Fig.\ref{F13}a-b. Furthermore the subsequent pulses interact with
periodically structured surface hence a grating-assisted
laser-surface coupling becomes effective \cite{Huang2009_1}
causing a decrease of the ripple wavelength. The experimental data
shows only minor modification of LIPSS periods that is
consistent with the hypothesis in Ref. \cite{Huang2009_1} that
because of strong thermal effect at the crater center, the
grating-assisted coupling is weak and the ripple wavelength is
unaffected by higher exposure, i.e. depends weakly on the
superimposed pulse number.

\section{Conclusion}

The influence of prompt and cumulative optical feedback contributions
in multi-shot fs-laser induced dynamics of surface ripples was investigated theoretically and experimentally.
The numerical simulations of periodic laser energy deposition on photo-excited diamond surface based on Sipe-Drude theory provided realistic
and detailed insight into microscopic mechanisms. The model identifies the impact ionization as a relevant process causing
optical breakdown of diamond in the trailing edge of the pulse resulting in plasmonically-active
substrate with negative bulk dielectric constant triggering the SPP-laser interference mechanism for surface ripple formation.
Fine ripples oriented perpendicularly to the laser polarization emerge for intensities in a narrow range above the optical breakdown threshold and
the transient increase of the carrier density above  this threshold results in the formation of near-wavelength surface ripples.
The interpulse feedback mechanisms involved in LIPSS formation are not considered by the present theory
and further work will be carried out in this direction. The obtained results lay the groundwork for utilizing diamond as a plasmonic material
supporting subwavelength and intense SPPs that is very promising for advanced optical applications.

\section*{Acknowledgements}
This work was partially supported by the Russian Foundation for Basic Research (projects nos. 17-52-53003, 17-02-00293 and 17-52-18023) and the grant of the RAS Presidium program,
 as well as by the Government of the Russian Federation (Grant 074-U01) through ITMO Visiting Professorship Program for S.I.K.
This material is based upon work supported by the Air Force Office of Scientific Research, Air Force Material Command, USAF under Award No. FA9550-15-1-0197 (T.A.).

\end{document}